\documentstyle[12pt,epsfig,epsf]{article} 
\newlength{\dinwidth}                       
\newlength{\dinmargin}                      
\newcommand{\xpom}{x_{_{I\!\!P}}}
\newcommand{\pom}{I\!\!P}
\begin{document}
 
\thispagestyle{empty}
\begin{flushleft}
{\tt hep-ph/9610nnn}
\end{flushleft}
\begin{flushright}
ISSN 0418-9833
\end{flushright}
\begin{flushleft}
DESY 96-218 \\
October 1996
\end{flushleft}

\vspace{1cm}
\begin{center}
{\Large \bf{Deep Inelastic Physics with H1}
\footnote{Invited talk held at the 
4th International Conference on Deep Inelastic Scattering, Rome, April 1996}}\\

\vspace{1cm}
\large{Max Klein} \\

\vspace{0.5cm}
DESY-Inst. f. Hochenergiephysik, Zeuthen \\
 Platanenallee 6, D-15738 Zeuthen,
  Germany. klein$@$ifh.de
\end{center}

\vspace{8cm}
\begin{abstract}
\noindent
  A summary is presented of deep inelastic
  physics results obtained by the H1 
  collaboration during its first three years of experimentation 
  at the electron-proton collider HERA.
\end{abstract}
\newpage
\section{Introduction}
About 20 years ago, at the time of the advent of  
asymptotically free field theories of strong interactions \cite{poli}, 
the option was seriously considered of a high energy 
colliding electron-proton machine
 CHEEP~\cite{CHEEP} at CERN and also at most of the other
 big accelerator laboratories~\cite{epo}.
It then seemed ``that hadrons contain quark-parton 
constitutents which are rather point-like and carry the weak and 
electromagnetic couplings of strongly interacting matter~\cite{qpm,feyn}.
 The nucleon 
seems to contain other constitutents without these couplings, which one 
may call 'gluons' without any prejudice as to their nature"~\cite{JE}.
The quark-gluon interactions explored with virtual photons of large mass
$Q^2 \geq M_p^2$ are the subject of present day deep inelastic 
physics. The Rome conference took place nearly 5 years after the first 
$ep$ collisions were succesfully observed by the H1 and ZEUS experiments at 
HERA.

 This report summarizes results reported to this conference
by the H1 experiment~\cite{H1}. The H1 Collaboration submitted 16 
papers which were discussed during the parallel sessions. For the 
introduction of the opening session a choice had to be made in order to
present a consistent review without  preempting the individual 
contributions. This choice emphasized the classical deep inelastic
physics subjects - structure functions, gluon distribution, $\alpha_s$, charged
currents and deep inelastic diffraction \cite{jed}- where
 H1 obtained important  new results, compare \cite{eis}.

The H1 Collaboration succeeded in presenting
 to the conference the first precision 
measurement~\cite{H1F} of the proton structure function  $F_2(x,Q^2)$ 
with an error as small as 5\% at $Q^2 \simeq$ 20~GeV$^2$, 
the measurement range extending  in Bjorken $x$ from
$3 \cdot 10^{-5}$ at  $Q^2$ = 1.5~GeV$^2$  to $x=0.32$ at  $Q^2$ = 
5000~GeV$^2$.  This result is discussed in sect.2.

A salient feature of the HERA experiments H1 and ZEUS is their ability to 
reconstruct the complete hadronic final state besides the scattered 
lepton. This offers many complementary ways for precision tests of QCD. 
 In particular,  as discussed in sect.3, information on
 the gluon distribution $xg(x,Q^2)$
can be obtained in a wide range of $x$ analyzing the 
observed scaling violations, the  longitudinal structure function $F_L$,
charm production $via$ photon-gluon fusion,
vector meson production and deep inelastic jet production. 

HERA will permit a very precise determination
of the strong coupling constant with a possible systematic error of $\delta 
\alpha_s(M_Z^2) \simeq 0.002$ \cite{kbp} once its ambitious running program
is approaching completion. Important steps towards this precision were 
presented to this conference, as discussed in sect.4,
by establishing the running of $\alpha_s(Q^2)$ in 
DIS jet production, by precisely describing the  $F_2$ scaling 
violations with a NLO DGLAP \cite{dglap} QCD procedure, and by 
determining $\alpha_s(Q^2)$ in the double logarithmic scaling 
approximation of the
low $x$  and large $Q^2$ behaviour of $F_2$.

Due to the high $Q^2$ range accessible, the HERA collider enables the 
study of neutrino physics and the electroweak interaction 
in the spacelike region. Sect.5 presents the first results of H1,
based on 130 charged current events with 6.4~pb$^{-1}$ luminosity,
probing the proton structure with virtual $W$ bosons at large $x$. 

Finally, sect.6 is devoted to hard diffraction events, those
as described in 1987 \cite{don} ``in 
which the target proton emerges isolated in rapidity" and which ``probe 
the quark and antiquark content of the pomeron, that is they measure the 
pomeron structure function". Based on the precision inclusive 
cross section data, important progress could be reported to this 
conference in the quantitative investigation of deep inelastic
diffraction.
%
\section{Measurement of  $F_2(x,Q^2)$}
%
%
Since the inclusive deep inelastic scattering  cross section
\begin{equation}
\hspace*{-0.4cm} \frac{d\sigma}{dxdQ^2} = \frac{2\pi \alpha^2}{Q^4 x}
        \cdot [(2(1-y)+y^2) F_2(x,Q^2) -y^2 F_L(x,Q^2)] = \kappa \cdot F_2
        \label{sigb}
\end{equation}  
depends on two variables only, the measurement of the scattered electron 
energy $E_e'$ and angle $\theta_e$, of the hadronic quantity 
$\Sigma_h=\Sigma_i(E_i-p_i^z)$ and of the hadronic angle $\theta_h$
gives rise to an overconstrained  
determination of the kinematics and permits maximum 
coverage of the available $(x,Q^2)$ range. Here $y$ is the inelasticity 
variable, $y=Q^2/sx$,  $s=4E_eE_p$ with the beam energies $E_p = 820$~GeV 
and $E_e=27.5$~GeV, and $\tan~\theta_h/2 = \Sigma_h/p_T^h$
 with $p_T^h=\sqrt{ (\Sigma_ip_x^i)^2  +(\Sigma_ip_x^i)^2}$,
 where the summation extends over all particles but the 
scattered electron.

The $F_2$ structure function data presented to this
conference~\cite{H1F,ub} was the last taken prior to the upgrade of the 
backward region of the H1 detector. 
  Compared to previous analyses a new level of accuracy 
was achieved  due to the larger statistics available
 for energy and angle calibrations and efficiency determinations. The 
measurement of $E_e'$ could be  calibrated to  an absolute 
scale accuracy of 1\% using the ``kinematic peak shape'' 
of the $E_e'$ distribution, a cross section enhancement for
 $E_e' \simeq E_e$ at $x=E_e/E_p 
\simeq 0.03$,  and the reconstruction of 
$E_e'=E(\theta_e, \theta_h)$  as functions of the two angles.
 The electron polar angle measurement was accurate to 1 
mrad and the hadronic energy scale was known to 4\%. An important error at 
lowest $Q^2$ was the uncertainty of the vertex reconstruction efficiency,
up to 8\% at large $x$ and 4\% at lowest $x$. 
The luminosity  of 
2.7~pb$^{-1}$ was measured  to 1.5\% accuracy using the energy 
spectrum of hard photons ($E_{\gamma} \geq 10$ GeV)  in the
reaction $ep \rightarrow ep\gamma$. 
\begin{figure}[b]\centering
\begin{picture}(280,240) 
\put(-58,-58){
\epsfig{file=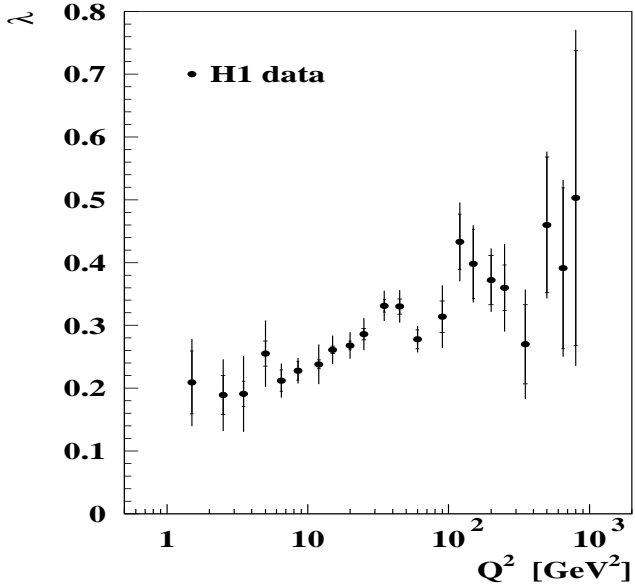,width=12cm,height=12cm,bbllx=0pt,bblly=0pt,bburx=557pt,bbury=792pt}}
\end{picture}
\caption{Derivative $\partial \ln F_2/ \partial \ln (1/x)$ determined
in each $Q^2$ bin of the H1 data
from a straight line fit to $\ln F_2=a+\lambda \ln (1/x)$ with the
full systematic error
correlations taken into account.
The inner error bar is the statistical 
error, the full error bar is the total error.}
\protect\label{xl}
\end{figure}
 For  the first time the H1 structure function 
measurement comprises data with a tagged, initial state radiated
 photon which enabled access to 
the lowest $Q^2$ region due to the reduction of the incoming energy $E_e$
by $E_{\gamma}$. The low $Q^2$ region was covered as well by data
taken with a $z$ vertex position shifted by +70~cm in proton beam
direction  which  extended the measurement of $F_2$ to  large 
values $\theta_e \leq 176.4^o$, $\theta$ being measured relatively to
the proton beam direction.  The accuracy at low 
$Q^2 \leq 8$~GeV$^2$ and at  large $Q^2 \geq 80$~GeV$^2$ is only about
10 to 20\% while in the region of the 
bulk data a systematic error down to 5\% was achieved. Although this data 
is more precise than the previous one, it should be kept in mind that 
real precision measurements of $F_2(x,Q^2)$ with errors of about 2-3\%
in  the whole kinematic range accessible by the HERA experiments
 are still to be performed~\cite{kbp}.  
%
%
%

 The H1 measurement extends the knowledge of $F_2$ from fixed
target lepton 
experiments  by about two orders of magnitude 
towards lower $x$.
A strong rise of $F_2$ is observed with decreasing $x$ 
at fixed $Q^2$. This rise is correctly reproduced by a NLO QCD fit, 
described in~\cite{H1F}, which starts from input gluon, sea and valence 
distributions at a chosen $Q_o^2$ value of 5 GeV$^2$ and generates that 
behaviour at larger and lower momentum transfers through the DGLAP 
evolution equations. The rise has been quantified by determining the 
exponent $\lambda$ of $F_2 \propto x^{-\lambda}$ at fixed $Q^2$ using only 
the H1 data, see fig.\ref{xl}.
 So far this slope is determined rather precisely only for the 
intermediate $Q^2$ range between about 8 and 80 GeV$^2$.
Note that there is a hidden possible $x$ dependence of 
$\lambda(Q^2)$ since with rising $Q^2$ the mean $x$ increases due to
the $y$ limitations of the kinematic region accessed, $\bar{Q}^2/\bar{x} \simeq
10^4$, roughly. If in the 
intermediate $Q^2$ range a common $x$ interval between $x=0.0008$ and 
$x=0.008$ is chosen the dependence of $\lambda$ on $Q^2$ gets  
slightly reduced. Higher precision data are needed, however, 
 to contrast this with  the expectation of a 
$1/\sqrt{ln(1/x)}$ dependence of $\lambda$ on $x$.

The $\lambda$ pa\-ra\-me\-ter not on\-ly quan\-ti\-fies the 
derivative  of $\ln F_2$ vs $\ln (1/x)$ 
 but is as well 
intimately connected with the behaviour of the 
photoproduction cross section because of $\sigma_{tot}(\gamma^* p) \propto 
F_2(W,Q^2)/Q^2$ where $W$ is the energy in the $\gamma^* p$ centre of
mass system, $W=\sqrt{Q^2/x}=\sqrt{sy}$ at low $x$. In the double
logarithmic scaling 
hypothesis \cite{ball} which goes back to the roots of QCD \cite{alva} the 
distributions at low $Q^2 < M_p^2$ are soft, $F_2 \propto x^0$,
 and $\lambda$ is expected to increase 
with $Q^2$, and to depend on $x$.  As was discussed at this conference,
the whole be\-ha\-vi\-our can be very well reproduced by an 
expression  $\lambda \propto \sqrt{\ln T/\ln (1/x)}$ with 
$T=\ln (Q^2/\Lambda^2)/\ln (Q_o^2/\Lambda^2)$ \cite{knd}.
In the alternative case of hard 
input or Pomeron distributions \cite{Y} one expects $\lambda$ for the 
singlet distribution, which essentially is $F_2$ at low $x$, not to depend 
on $Q^2$ above some threshold value around 12~GeV$^2$. While the
data seem to favour a  $Q^2$ 
dependence of $\lambda$, reasonable fits have been obtained with
a NLO factorization ansatz \cite{Y}. Improved precision at all $Q^2$
will finally settle 
\begin{figure}[t]\centering
\begin{picture}(180,260) %
\put(-200,-170){\epsfig{file=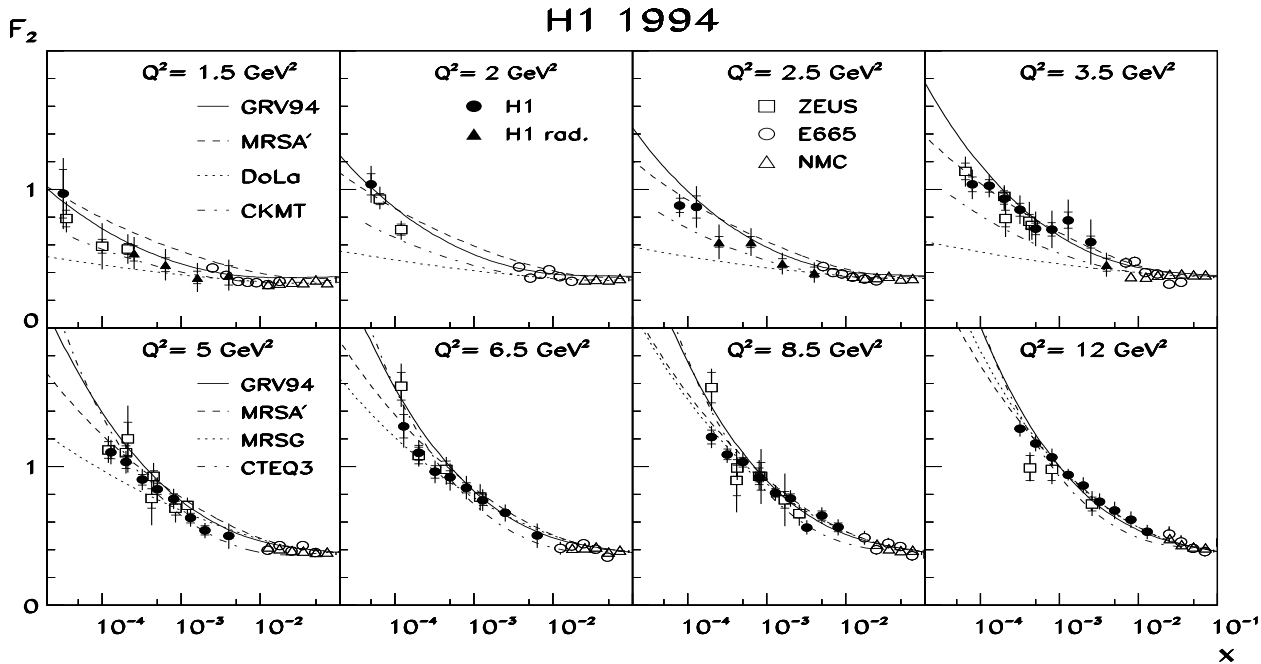,bbllx=0pt,bblly=0pt,bburx=557pt,bbury=792pt,width=18.5cm,height=19.4cm}}
\end{picture}
        \caption{Measurement of $F_2$ at low $Q^2$ compared with ZEUS,
          E665 and NMC data and various structure function parametrizations.}
        \protect\label{f2mo}
\end{figure} 
this question which is an example of how intimately 
deep inelastic physics is 
related to its transition to the photoproduction region~\cite{levy}.

 At low $Q^2$ the structure function data have been compared with various 
parametrizations of $F_2$, see fig.\ref{f2mo}. There are three 
observations to be emphasized: i) the data disfavour Regge based models 
for $Q^2$ larger than $M_p^2$; ii) the dynamical parton distributions of 
GRV \cite{GRV} agree generally well with the measurement but
have the tendency of overshooting the data at the lowest 
$x$ values which may be cured by adjusting the starting point of the 
evolution and of $\Lambda_{QCD}$. The GRV approach is based on 
soft, valence like input distributions and thus fits to the double 
scaling concept; iii) various sets of global fits from  CTEQ \cite{CTEQ}
and MRS \cite{MRS}, which use the 1993 or already the 1994 data, 
reproduce the behaviour of $F_2$. Of particular interest are  the starting 
distributions which when considered at $Q^2 \simeq M_p^2$ seem to favour 
a soft singlet distribution as well. 

  The H1 collaboration has an ongoing 
long term programme for measuring $F_2$  with higher 
luminosity and an upgraded detector which should enable us to study
the nature of quark-gluon interactions at high parton densities with
$F_2$ much 
more precisely than so far. In particular, measurements with greater
precision at lowest accessible $x$ values in the DIS region may yet reveal BFKL 
interaction dynamics~\cite{bfkl}. The prediction of the low
$x$ behaviour of $F_2$ according to BFKL still awaits the NLO
calculations to be completed. Phenomenologically a unified treatment
of DGLAP and BFKL dynamics is rather succesfull~\cite{KMS}.
Interesting suggestions have been presented to search for BFKL
dynamics in forward jet production~\cite{alb} and $E_T$
spectra~\cite{kuhl} but no conclusive evidence has been established
yet with the H1 data analyzed so far.
\section{Access to the Gluon Distribution $xg(x,Q^2)$}
\subsection{Scaling Violations}
 Scaling is violated in any interacting field theory \cite{col}, logarithmically in 
 QCD.
\begin{figure}[t]\centering
\begin{picture}(180,380) 
\put(-130,-70){\epsfig{file=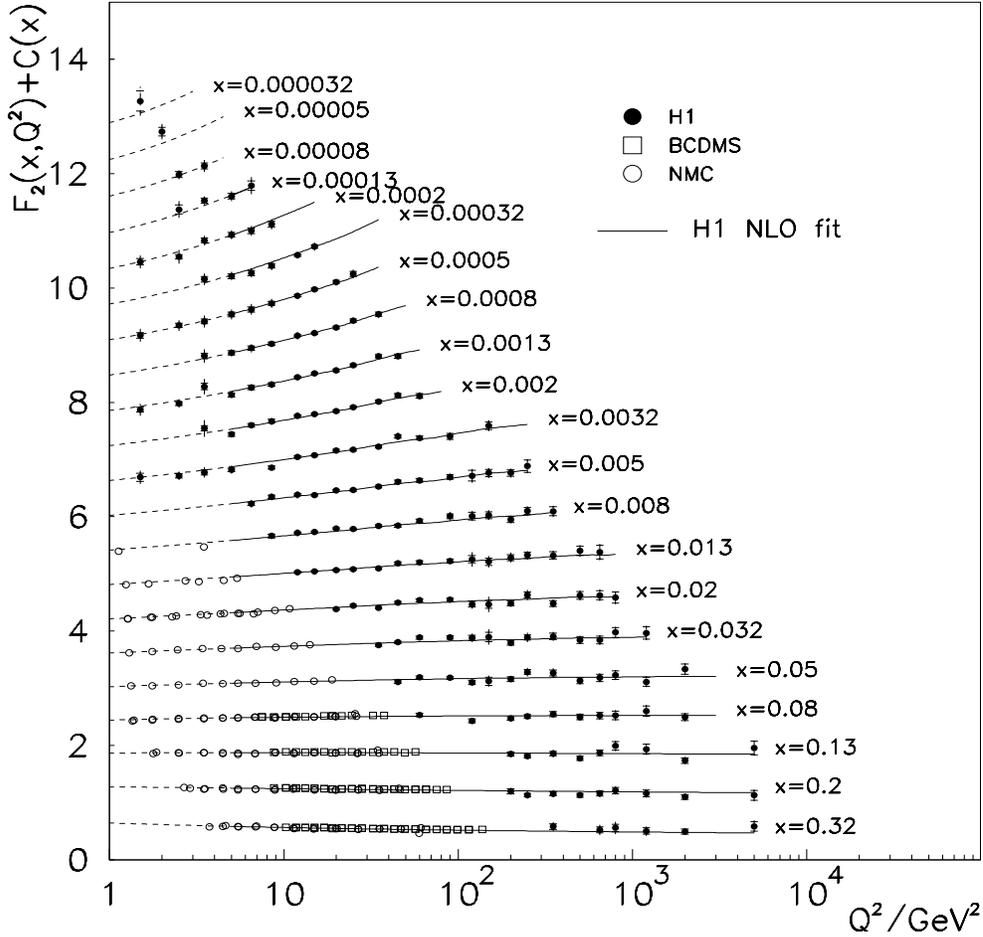,bbllx=0pt,bblly=0pt,bburx=557pt,bbury=792pt,width=13.6cm,height=18cm}}
\end{picture}
\caption{Dependence of $F_2(x,Q^2)$ on $\log Q^2$ as measured by the H1 
  experiment. The data is well described by a NLO QCD fit and smoothly 
  extends the fixed target measurements into the higher $Q^2$ region
  albeit with less precision so far.}
\protect\label{f2q}
\end{figure}
 The pattern of scaling violations has long ago been anticipated. 
 At low $x$ the structure function was predicted to increase
  with $Q^2$ due to quark pair 
 production in the gluon field while at high $x$ it should fall because of gluon 
 radiation from quarks. Due to momentum conservation the integral of 
 $F_2$ is about independent of $Q^2$ which leads to a turn over point at 
 $x \simeq 0.13$ where scale invariance holds. 
 This pattern is observed by H1 over 
 a large range of $x$ and $Q^2$ as can be seen in fig.\ref{f2q}. The  scaling violations  for $x \leq 0.01$ determine directly the 
 gluon distribution. In  leading order perturbation theory
 one has approximately \cite{prytz}
\begin{eqnarray}
  \frac{\partial F_2(x/2, Q^2)}{\partial \ln Q^2} & \simeq &
  \frac{10}{27} \frac{\alpha_s(Q^2)}{\pi} \cdot xg(x,Q^2)
  \label{sig}
\end{eqnarray}
 because the quark contribution to 
 the derivative of $F_2$ is  small ($\leq 10\%$) for $x < 0.01$. 

The  measurement of scaling violations is the most precise way to 
 access the gluon distribution at low $x$. Fig.4 shows the result 
 of a NLO QCD fit performed by H1~\cite{H1F} using only $F_2$
 structure function 
 data. In this fit  all parameters describing the singlet, 
 gluon and valence quark distributions were fitted, the correlation of 
 experimental systematic errors was considered and the QCD scale parameter 
 $\Lambda_{QCD}$ was fixed to 263~MeV. The gluon distribution rises 
 towards low $x$ the more the greater is $Q^2$. Such a behaviour is 
 a result of the interaction dynamics as inherent in the DGLAP evolution 
 equations. So far no deviation from this concept has been experimentally 
 found in the behaviour of $F_2$ for $Q^2 \geq M_p^2$ at low $x$. Thus one may 
 determine the gluon distribution at low $x$
 with this method down to $Q^2$ values 
 around 1~GeV$^2$. This rather low $Q^2$ value phenomenologically may not 
 be surprising: at low $x$ the DGLAP equation for the gluon can be 
 exactly solved \cite{muel,Y} leading to the behaviour
\begin{eqnarray}
\vspace*{0.4cm}
   xg(x,Q^2)  \propto  \exp
\sqrt{c \cdot \ln \frac{\ln (Q^2/\Lambda ^2)}{\ln (Q_o^2/\Lambda^2)} \cdot \ln (1/x)}.
\label{gmuel}
\end{eqnarray}
 This expression for  $Q^2 \rightarrow Q^2_o$  degenerates 
to a normalization constant but it is
 still evolving with $x$ and $Q^2$ even below 
 1~GeV$^2$ for $Q^2_o$ around 0.4~GeV$^2$, which is the GRV starting
 scale parameter 
 and the value one can obtain by fitting a double logarithmic expression 
 to the $F_2$ data~\cite{knd}.
\begin{figure}[!ht]\centering
\begin{picture}(180,240) 
\put(-60,-50){\epsfig{file=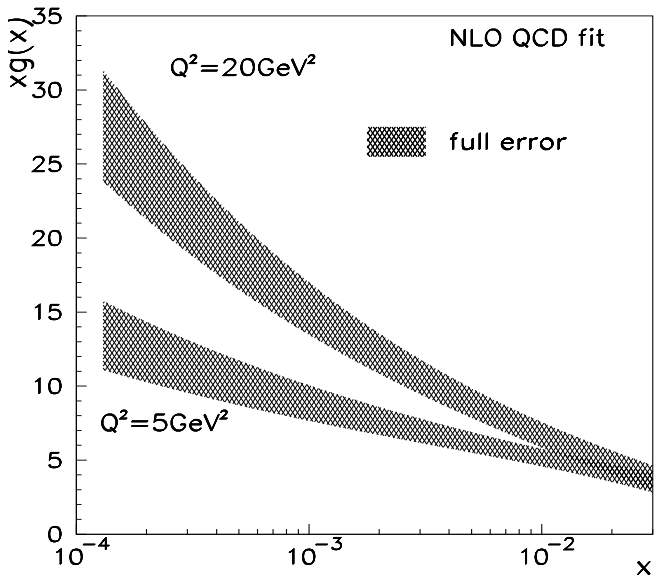,bbllx=0pt,bblly=0pt,bburx=557pt,bbury=792pt,width=9cm,height=11cm}}
\end{picture}
        \caption{Gluon distribution determined by H1 from a NLO QCD fit to the 
        H1, NMC and BCDMS proton and deuterium structure function data.}
        \protect\label{glu}
\end{figure}
%
 \subsection{Longitudinal Structure Function $F_L(x,Q^2)$}  
In the quark-parton model the longitudinal structure function is zero for 
spin half quark-photon scattering \cite{calg}. In Quantum 
Chromodynamics $F_L$ acquires a non zero value because of gluon radiation. 
At low $x$, as for \mbox{$\partial F_2 / \partial \ln Q^2$}, the 
gluon distribution dominates which allows  access to $xg$ via the
relation \cite{AM}  
\begin{equation}
        F_L = \frac{\alpha_s}{4 \pi} x^2 
        \int \frac{dz}{z^3} \cdot [ \frac{16}{3}F_2 + 8 
        \sum Q_q^2 (1-\frac{x}{z}) g] 
        \label{AM}
\end{equation}
with the quark charges $Q_q$ and 
written for four flavours, see \cite{GRV} for the treatment of charm.  The 
longitudinal structure function, together with $F_2$, allows an important 
consistency test of QCD, as eq.\ref{AM} illustrates: at low $x$ 
the gluon distribution determines simultaneously
the scaling violations of $F_2$  and 
the size of $F_L$. $F_L$ may differ by a factor of  two in the BFKL approach
 \cite{sal,bal} from the DGLAP prediction and the predictions 
based on the factorization
ansatz and on the double scaling hypothesis differ~\cite{Y}.  A 
measurement of $F_L$  to about 10\% precision is 
necessary to determine  the behaviour of $F_2$ at the lowest  
$x$ values which are accessible at high $y=Q^2/sx$. 
The H1 collaboration 
has presented a determination of $F_L$ at low $x$ to this conference
for the first time~\cite{mk}. 
\begin{figure}[h]\centering
\begin{picture}(160,220)
\put(-80,-100){\epsfig{file=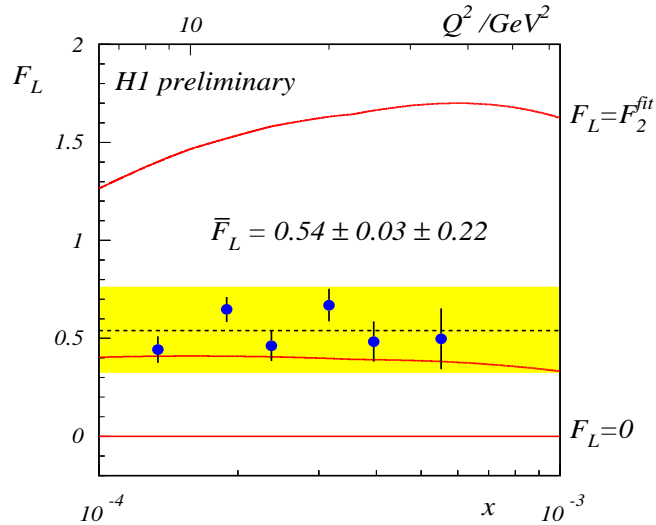,bbllx=0pt,bblly=0pt,bburx=557pt,bbury=792pt,width=10cm,height=13.3cm}}
\end{picture}
\caption{Measured longitudinal structure function $F_L$ for $y=0.7$ as 
functions of $Q^2$ and of $x=Q^2/sy$. The error bars are the statistical 
error, the band is the systematic error common to all points. The curve 
        is a NLO QCD calculation of $F_L$ using the gluon and quark 
        distributions determined  from the lower $y$ H1 and BCDMS data.}
        \protect\label{FL}
\end{figure} 
At high $y \geq 0.6$ the $y$ dependent weight factors of $F_2$ and $F_L$
in eq.\ref{sigb}
become of comparable size. Therefore the usu\-al technique of extracting 
$F_2$ with a calculated $F_L$ was reversed  and $F_L$ was  determined
after subtracting the $F_2$ contribution to the measured cross section.
Contrary to the $F_2$ extraction which is based on {\it ad~hoc} calculated
values 
for $F_L$, this determination of $F_L$ utilized the cross
section measurement from a different $y$ region which was extrapolated to 
high $y$ in order to subtract the $F_2$ contribution. This extrapolation
used a NLO QCD fit to $F_2$ performed in the lower $y$ region. 

The $F_L$ measurement was performed using the most precise part of
the 1994 data, for 
$7.5 < Q^2 < 42~$GeV$^2$, with a luminosity of 1.25~pb$^{-1}$. Access to 
$F_L$ required  the $y$ range to be extended from a maximum of 0.6 \cite{H1F} 
to 0.78, given by a reconstruction limit of the 1994 data placed near
$E_e'=6~$GeV.
This was achieved using a lower energy calorimetric trigger combined
with a central track trigger.
A complete cross section
reanalysis had to be performed which
 is in very good agreement with the published analysis. 

The resulting longitudinal structure function is given in
fig.\ref{FL}. It has been determined at six different $x$ or $Q^2$
values for a  common 
$y$ value of 0.7. An average, preliminary value of $F_L=0.54 \pm 
0.03 (stat) \pm 0.22 (syst)$ has been determined at an average $Q^2$ 
value of 18 GeV$^2$ and $x=0.0003$. The systematic error 
 is mainly depending on $y$ only. It 
includes both the measurement errors of the  cross section at high $y$ and 
the uncertainty of the subtracted $F_2$. That is due to  the $F_2$ data errors 
at lower $y$, which partially get compensated, 
and to the fit procedure. On average the   
measured $F_L$ value is 2.6 standard deviations
 larger than zero and 4 to 5 standard deviations away from 
$F_2$. With the specific values of $F_2$ 
and $F_L$ as indicated in fig.\ref{FL}, $R=F_L/(F_2-F_L)$
 is about 0.5 with errors roughly 1.5 
times larger than those of $F_L$.  Compared to fixed target measurements at larger $x$ 
values~\cite{mil}, $R$ is rather large which is consistent with the gluon 
distribution at low $x$

The $F_L$ extraction procedure assumes  
that $F_2$  follows NLO QCD. This 
is a reasonable assumption given the 
good agreement with QCD over many orders of magnitude in $Q^2$. Yet, it 
may not necessarily be true because one is exploring here at each
$Q^2$ the lowest 
accessible $x$ values where $F_2$ might depart from the expected 
behaviour. A measurement free of this assumption requires the 
proton beam energy to be lowered
which is an option planned in the HERA programme. Due to the reduced
energy, however,  this measurement  will determine $F_L$ 
values at about two times larger $x$ at a given $Q^2$ than reachable 
with the subtraction method. At these higher $x$, for
$Q^2 \geq 10~$GeV$^2$, the QCD description of $F_2$ is certainly valid.
Yet, the comparison of the subtraction method with the result from two
energy measurements will be very interesting for minimizing the systematic
uncertainty of the $F_L$ determination. At lower
$Q^2$ the $F_2$ QCD extrapolation becomes questionable and the two
energy measurements shall be essential in determining $F_L$.
\subsection{Charm Structure Function}
To leading order the charm structure function $F_2^c$ is a direct 
measure of the gluon distribution because of the relation 
\begin{eqnarray}
  F_2^c & = & \frac{Q^2 \alpha_s(\mu^2) Q^2_c}{4 \pi^2 m_c^2}
      \cdot \int_{x(1+4m_c^2/Q^2)}^{1} \frac{dz}{z} \cdot 
      zg(z,\mu^2)~C^{(0)}
\label{f2c}
\end{eqnarray}
with the scale parameter $\mu$, the charm quark charge $Q_c$ and 
mass $m_c$ and the lowest order coefficient function $C^{(0)}$. 
At low $x$ the quark 
contribution to the higher order expression \cite{riem} is small and the 
dependence of $F_2^c$ on the renormalization and factorization scale is 
$ \leq 10$\% \cite{dursf}. The measurement of the charm structure function allows an
almost local determination of $xg$ because at a given $x$ the largest 
part of the integral stems from a narrow $z$ interval.
\begin{figure}[ht]\centering
\begin{picture}(160,180) 
\put(-120,-70){\epsfig{file=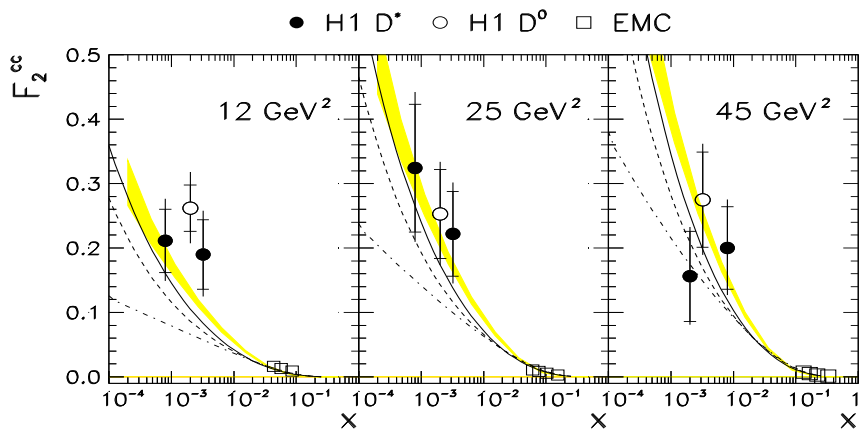,bbllx=0pt,bblly=0pt,bburx=557pt,bbury=792pt,width=13.cm,height=15.2cm}}
\end{picture}
        \caption{Measurement of $F_2^c$ based
         on the observation of $D^{*+}$ and 
        $D^0$ production in deep inelastic $ep$
        scattering at low $x$. The full 
        error bar represents the total error which is dominated by statistics 
        (inner bars).}
        \protect\label{fc}
\end{figure}
 The H1 collaboration has presented preliminary data on $F_2^c$ 
to this workshop \cite{kdr}.
The extraction of $F_2^c$ relies on the observation of open charm 
production in deep inelastic scattering. Based on an integrated
luminosity of 2.97~pb$^{-1}$ 144 $D^0$ and 103 $D^*$ events were 
found after statistical background subtraction in the $K \pi$ and the 
mass difference distribution, $m(D^*)-m(D^0)$, respectively. Taking into 
account the fragmentation function of $c \rightarrow D$ the $D$ meson 
cross sections were determined and $F_2^c$ was derived independently 
from both $D$ meson data samples. The observed 
transverse momentum spectra of inclusive $D^0$ and $D^*$ production were 
found to be in good agreement with photon-gluon fusion calculations. 
The measured $F_2^c$ is shown in fig.\ref{fc} together with 
different NLO calculations of $F_2^c$. Remarkable agreement is
observed between the measurement and the H1 calculation of $F_2^c$
using essentially $xg$ from the QCD fit to $F_2$ (dashed area in
fig.\ref{fc}). The measurement extends the EMC 
data by two orders of magnitude down to  $x \geq 0.0008$. Within the 
errors $F_2^c$ does not depend on $Q^2$ and the ratio of $F_2^c$ to $F_2$ 
is $0.237 \pm 0.021~^{+ 0.043}_{- 0.039}$ at mean values of $Q^2
\simeq 26$~GeV$^2$ and $x \simeq 0.002$.
A sizeable charm contribution to $F_2$ at low $x$ could be 
expected from eq.\ref{f2c} because of the large gluon distribution.
 
The data presented are the first results on $F_2^c$ from H1. With 
increased luminosity the charm structure funct\-ion and the extraction of 
the gluon distribution will be of increasing importance with an ex\-pec\-ted 
precision of about 10\% \cite{kd}.\\
\subsection{Elastic $J/\psi$ Production}
Results were presented by H1 on the deep inelastic and photoproduction of 
the vector mesons $\rho,~\omega,~\Phi,~\rho'$ and $J/\psi$
based on data taken in 1994 with a luminosity of about 3~pb$^{-1}$. These 
provide an impressive amount of detailed information on the 
energy dependence and relative size of production cross sections, 
angular distributions and $t$ dependence \cite{cler}. The  $J/\psi$
and $\rho$ production have been selected here which provide some information on 
the gluon distribution.
 
Elastic $J/\psi$ production has been viewed as proceeding via diffractive 
scattering \cite{lan} with a rather mild dependence of the cross section 
on $W$,  the energy in the $\gamma p$ centre of mass system. In 
perturbative QCD,
in leading order, the cross section for elastic $J/\psi$ production is 
given by \cite{rys,brod}
\begin{equation}
\vspace*{-0.2cm}
  \frac{d\sigma}{dt}  =
  \frac{\Gamma_{ee} m_{\psi}^3 \pi^3} {48 \alpha \mu^8} \cdot 
  [\alpha_s(\mu^2) \cdot xg(x,\mu^2)]^2
        \label{jei}
\end{equation}
with $\mu^2=m_{\psi}^2/4$ and $x=m_{\psi}^2/W^2$, i.e. it is a direct 
measure of the gluon distribution with $m_{\psi}$ providing the hard 
        scale \cite{frank}. The energy dependence of the $J/\psi$
production cross section is related to the low $x$ behaviour of the gluon 
density and expected to be rather steep.
\begin{figure}[b]\centering
\begin{picture}(160,180)
\put(-50,-30){\epsfig{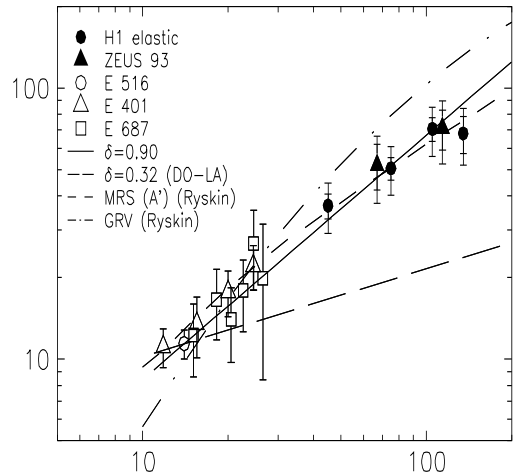}}
\end{picture}
        \caption{Total cross section for elastic
         $J/\psi$ photoproduction versus 
        the cms energy of the $\gamma p$ system.
        The inner error bars of the H1 
        data points are the statistical error,
        the full error bar represents the 
        total error.}
        \protect\label{JW}
\end{figure}
  Elastic $J/\psi$ events have been analyzed \cite{cler,H1J}
 in four intervals of $W$ in 
$J/\psi \rightarrow \mu^+ \mu^-$  with 2.7~pb$^{-1}$ (233 events) 
and in  $J/\psi \rightarrow e^+ e^-$ with 2.0~pb$^{-1}$(165 events) using 
a set of five different triggers sensitive to charged lepton production. 
Nearly background free $J/\psi$ samples were selected with a 
reconstructed mass of $3.10 \pm 0.01$~GeV ($\mu \mu$) and  
$3.08 \pm 0.02$~GeV ($e e$) to be compared with the PDG value of 
$m_{\psi} = 3.097$~GeV. Fig.\ref{JW} illustrates the $W$ 
dependence of the measured cross section. A parametrization of 
$ \sigma\, \propto\, W^{\delta}$ clearly favours 
a strong $W$ dependence with $\delta \simeq 0.9$ over the diffractive 
hypothesis with $ \delta $ between  0.22 and 0.32. This trend is observed 
with the large  $J/\psi$ production cross section measured
 at $W$ around 100~GeV as compared to previous fixed target 
experiments  but it is as well consistent with the $W$ dependence 
established by the H1 data alone. The energy dependence of the
cross section can be well described  by the model \cite{rys} with the 
MRSA' gluon parametrization which 
 behaves like $x^{-0.2}$ at low $x$.

Data were presented also on inelastic $J/\psi$ photoproduction
 \cite{cler,H1J}, about 85 events with a $J/\psi$ decay into $\mu^+ \mu^-$,
which are in good agreement with NLO QCD calculations \cite{krae}
both in normalization and energy dependence. Yet, reliable predictions 
were obtained only for inelasticity values 
$z=y_{\psi}/y \leq 0.8$ and transverse momenta
 $P_T^{\psi} \geq 1$~GeV. This cuts into 
the low $x$ region reducing 
the data by about one half and thereby the
sensitivity of the cross section to 
the still possible variations of the gluon density.  Nevertheless, this 
process may 
provide a rather precise determination of $xg$ but unlikely at the smallest $x$  
accessible by the data.
\subsection{Elastic $\rho^o$ Production}
Deep inelastic $\rho$ production is another process sensitive to the 
gluon distribution  with a cross section for longitudinally polarized 
photons of 
\begin{equation}
  \frac{d\sigma}{dt} \propto
  \frac{[\alpha_s(\mu^2) \cdot xg(x,\mu^2)]^2}{Q^6} \cdot C_{\rho}.
        \label{rho}
\end{equation}
The situation is similar to $J/\psi$ production with a predicted weak $W$ 
dependence in diffractive models \cite{lan} and a stronger rise expected 
in QCD. Yet, for $\rho$ (and $\phi$)
 production it requires a large virtuality of the 
process to introduce a hard scale. Another expectation from eq.\ref{rho} 
is a $Q^2$ dependence of $Q^{-2n}$ with $n \simeq 2.5$ since at low $x$ 
$(\alpha_s xg)^2 \propto \sqrt{Q^2}$ and $C_{\rho}$ only weakly depends 
on $Q^2$ \cite{koe}. 
%
\begin{figure}[t]\centering
\begin{picture}(160,240)
\put(-54,-1){\epsfig{file=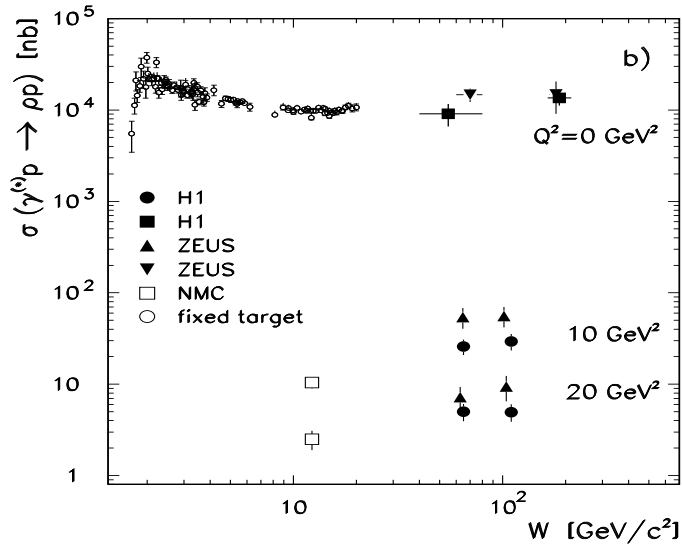,width=10.cm,height=8cm}}
\end{picture}
        \caption{Total cross section for elastic $\rho$ production versus 
        the cms energy of the $\gamma p$ system compared with ZEUS and fixed 
        target data.}
        \protect\label{rhof}
\end{figure}
  Fig.\ref{rhof} summarizes the elastic 
 cross section measurements of $\rho$ 
photoproduction \cite{H1rp} and large $Q^2$  production \cite{H1re}. The 
$\rho$ decay pions were reconstructed in the central drift chambers apart 
from photoproduction data  at $W \simeq 200$~GeV which had to use
 a calorimetric 
measurement because of the large boost of the $\rho^o$ rest frame which 
forced the pions outside the tracker acceptance region. At $Q^2 \simeq$ 
10~GeV$^2$ 104 events were analyzed and at $Q^2 \simeq$ 20~GeV$^2$ 78 
events from a luminosity of 2.8 pb$^{-1}$. The result is somewhat lower
than the ZEUS data which have a 31\% normalization
uncertainty \cite{ZEUSr} not drawn in
fig.\ref{rhof}. 
A combination of the H1 
results with the NMC data reveals an increase of the cross 
section  $\propto W^{\delta}$ with $\delta=0.56 \pm 0.20$ at 
$Q^2=10~$GeV$^2$ and $\delta=0.40 \pm 0.24$ at $Q^2=20~$GeV$^2$. The 
error includes the uncertainties of both experiments. The photoproduction 
cross section, however, does not depend strongly on $W$ which differs 
from the $J/\psi$ result. The $Q^2$ dependence has been measured and 
$n=2.5 \pm 0.5 \pm0.2$ obtained with a dominant statistical error. 
These results support the QCD approach to deep inelastic
$\rho$ production and are consistent with expectations from the behaviour 
of the gluon distribution at low $x$. It remains to be seen whether 
future more precise data and the quantitative understanding of 
non-perturbative effects \cite{koe,nonp} lead to a consistent picture of this 
interesting process.    

\subsection{Jets in Deep Inelastic Scattering}
While gluons had not been `seen' yet, the CHEEP proposal \cite{CHEEP} 
anticipated the existence of three jet events in deep inelastic 
scattering with a cross section written as the sum of photon-gluon
fusion
($\sigma_{pgf}$) 
and  Compton scattering  ($\sigma_{com}$) contributions \cite{geor}  as
\begin{equation}
\vspace*{-0.2cm}
  \sigma_{jet} = \alpha_s(Q^2) \int \frac{dx}{x} 
    [xg(x,Q^2) \cdot \sigma_{pgf} + 
    \sum_{q} xq(x,Q^2) \cdot \sigma_{com}].
        \label{gjet}
\end{equation}
Here three jets actually denote the proton remnant centered around the 
beam line and the two quark jets or a quark and a gluon jet.
To leading order one has calculated and
 corrected for the Compton contribution, 
identified $ x$ with 
 $x_{Bj}(1+m^2/Q^2)$, $m$ being the two jet mass, and unfolded 
directly the gluon distribution with some assumption on $\alpha_s$ 
\cite{H1lg}. The H1 Collaboration has presented a NLO extension of
this approach to this conference~\cite{kon} which required
 to replace the direct unfolding by  a fitting 
procedure  solving the cross section integrals with a Mellin 
transformation technique \cite{grau}.
%
\begin{figure}[htb]\centering
\begin{picture}(160,220) 
\put(-60,180){\epsfig{file=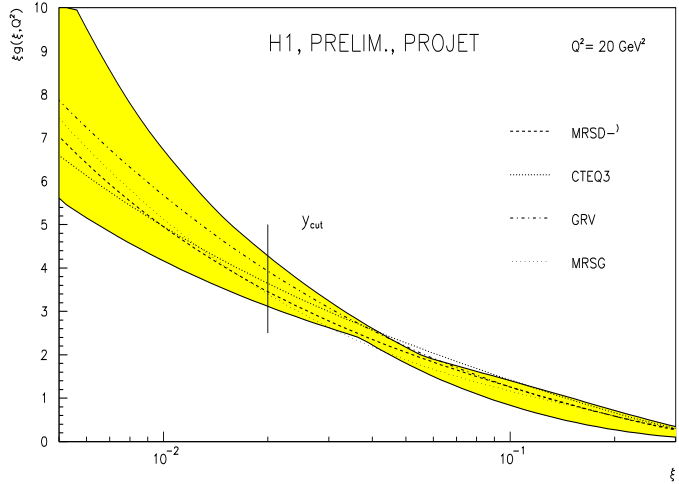,width=6.5cm,height=9.cm,angle=-90.}}
\end{picture}
        \caption{The gluon density obtained from DIS three jet production 
        compared to standard gluon parametrizations based
         on structure function 
        and direct photon data. The band represents the analysis uncertainty.}
        \protect\label{fglj}
\end{figure}
 Four values of the measured relative rate of three jet 
events for $Q^2$ between 40 and 4000~GeV$^2$ were used as observables.
 The fit procedure 
resembles the structure function QCD analysis as an assumption  
 $xg = a x^b\cdot$ $(1-x)^c (1+dx)$ was made at $Q_o^2=4$~GeV$^2$ 
 and the gluon distribution obtained through the QCD evolution at 
 higher $Q^2$. Fig.\ref{fglj} represents the first gluon distribution 
 determined by H1 from DIS jet production analyzed in NLO for $ x \geq 
 y_c$ where $y_c$ is the minimum resolution parameter in the JADE
 jet finding 
 algorithm applied here \cite{JADE}. The  jet formation was simulated
 to NLO using the PROJET program \cite{PROJ} and the
 transition of the observation 
 from the jet to the parton level was done using LEPTO \cite{lept}.
  The result is in good agreement with the standard 
 gluon parametrizations and the method promises for a more accurate 
 gluon determination at HERA in the deep inelastic regime for $x > y_c
 \simeq 0.02$.
  
\section{The Strong Coupling Constant $\alpha_s(Q^2)$}
\subsection{$\alpha_s(Q^2)$ from Jets}
The relative three jet rate $R_{2+1}$ has long been viewed as a way to
measure $\alpha_s(Q^2)$. A theoretical and practical complication of this 
$\alpha_s$ determination is due to the fact that hadrons are observed 
instead of partons and that the jet definition remains to be ambiguous to some 
extent.  Shortly before the Rome conference a new  NLO 
jet simulation program, MEPJET~\cite{mepj}, became available
 which allows tuning the program to any of 
the different procedures of jet recombination. This is promising as it 
ensures a very similar treatment of data and simulation. The H1 
Collaboration presented a progress report~\cite{kon} on the determination of
 $\alpha_s(M_Z^2)$ for different 
recombination schemes using the PROJET and the MEPJET code.
In MEPJET there is a 
reduced dependence of the $\alpha_s$ values on the recombination algorithm.
The central 
value in the `E' scheme comes out to be around 0.112 instead 
of 0.127 using PROJET. This is mainly due to a decrease of the measured 
$\alpha_s$ value at the lowest $Q^2 \simeq 200~$GeV$^2$ where the neglect of 
terms $\propto y_cW^2$ in PROJET becomes relevant. 

At the present stage of the analysis
this $\alpha_s$ determination has  systematic errors of about 
0.010 \cite{jaff} due to a hadronic energy  measurement scale error of 4\%, to 
hadronization  and parton density uncertainties, statistics, scale and 
$y_c$ effects. Further understanding of the whole procedure and an 
increase of the luminosity giving access to the highest $Q^2$ region 
should permit a substantial reduction in this uncertainty.

\subsection{QCD Analysis of Scaling Violations}
The classical method to determine the strong coupling constant in deep 
inelastic scattering has been to investigate the scaling violations of 
$F_2(x,Q^2)$ which lead to $\alpha_s(M_z^2) = 0.113 \pm 0.005$
\cite{marc} based on the BCDMS and SLAC experiments.
The distinction between the gluon distribution and 
$\alpha_s$ is practically difficult because whenever $xg$ appears in
the expression for a cross section it is naturally multiplied with  
$\alpha_s$ which, however, has a unique $Q^2$ dependence. The BCDMS/SLAC value was
determined using data at rather large $x$ 
where the contribution of the gluon distribution to the QCD $F_2$ 
evolution equation is small. Loosely speaking it is a value 
determined from gluon bremsstrahlung rather than quark pair production 
from gluons. The latter dominates at HERA. 

The H1 Collaboration has not yet presented an $\alpha_s$ determination 
based on the NLO QCD description of its structure function
data. There is sensitivity but 
$\Lambda_{QCD}$ is correlated to some of the many fit parameters
and improved precision of the $F_2$ data in the 
full kinematic range is desirable. In \cite{bfa} $\alpha_s$ 
was determined from the 1993 H1 data  by fixing the high $x$ 
parameters using global fits. A central value of 0.120 was obtained 
which is also somewhat favoured over 0.113 in a recent global analysis 
including the Fermilab jet data~\cite{MRS}. 

The H1 Collaboration has 
shown that double logarithmic scaling \cite{bfa} in combinations 
of the variables
$\ln (x_o/x)$ and $\ln (\alpha_s(Q_0^2)/\alpha_s(Q^2))$ 
holds to some 
approximation \cite{H1F}. This assumption has been used 
and a reanalysis was made which determined 
$\alpha_s(M_Z^2)$ to $0.113 \pm 0.002 \pm 0.006$ \cite{knd} using the
1994 $F_2$ H1 data only.
The advantage and the difficulty of this approach is that  here $xg$ and 
$\alpha_s$ appear to be decoupled: the gluon distribution is not 
entering anymore as it is essentially determined through eq.\ref{gmuel}.
This approach perhaps will permit a rather precise $\alpha_s$ 
determination if double logarithmic scaling continues to be 
supported by $F_2$ 
precision data and if the theoretical approximations can be better understood. 

The $\alpha_s$ determinations have 
a remarkable theoretical uncertainty due to factorization and  
renormalization scale uncertainties which were reconsidered
recently~\cite{rbv}. NNLO calculations of the splitting functions 
seem to be unavoidable for matching the envisaged experimental 
precision of $\alpha_s(M_Z^2)$ at HERA of $\simeq 0.002$ \cite{kbp}.

\section{Weak Charged Currents}
 Charged current events in H1 have the spectacular signature
 of a hadronic system of large transverse momentum, 
$p_T^2=Q^2 (1-y)$, which remains unbalanced as the (anti)neutrino escapes 
detection. 
Based on a luminosity of 2.7~pb$^{-1}$ for 1994 and 
3.7~pb$^{-1}$ for 1995 25 $e^-$ and 105 $e^+$ induced charged current 
events were measured with $p_T > 25~$GeV. The kinematics are
reconstructed using the $\Sigma$ method~\cite{bb}. Background is efficiently
removed with vertex, event topology
and calorimeter timing requirements. Details of the analysis are 
described in \cite{ccan}. Despite the limited statistics 
three basic observations were made: i) at $Q^2 \simeq G_F/\sqrt{2}e^2$
 the neutral and charged 
current inclusive cross sections become of similar size, see \cite{ccan}; 
ii) since HERA is equivalent to a 50 TeV neutrino beam 
fixed target experiment, H1 was able to discover departures from the 
linear energy dependence of the cross section with a measured propagator 
mass of $m_W = 84^{~+9~+5}_{~-6~-4}$ GeV in agreement with the most 
precise $W$ mass measurements at the Tevatron; iii) the $W$ boson 
can be used to probe the proton structure at very large $Q^2$. A first 
measurement of the $x$ distribution in charged current positron 
scattering, still not bin size corrected, is shown in fig.10. 
\begin{figure}[h]\centering
\begin{picture}(160,190)
\put(-90,-120){\epsfig{file=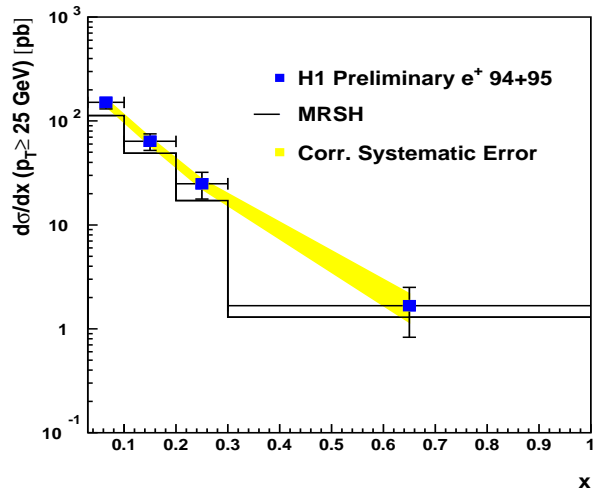,bbllx=0pt,bblly=0pt,bburx=557pt,bbury=792pt,width=10cm,height=12cm}}
\end{picture}
        \caption{Weak charged current cross sections measured by H1 as 
        a function of $x$.}
        \protect\label{cccx}
\end{figure}
The result is somewhat higher than 
the calculated cross section extrapolating the MRSH distributions to the 
high $Q^2$ region. Higher luminosity will permit  
access to the valence quark region and to a measurement of the up and 
down quark proton contents. As has been demonstrated 10 years ago 
\cite{bkn}, HERA 
may permit the determination of the electroweak mixing angle sin$^2 \theta$ with a 
precision of about 0.002 using the neutral to charged current cross 
section ratio in $e^-$ scattering.
This will be an interesting consistency test of the 
electroweak theory performed in the spacelike region.\\

\section{Deep Inelastic Diffraction}
A new experimental and QCD analysis of diffractive deep inelastic
scattering was presented by H1~\cite{paul} based on about 20,000 events obtained for
$2.5 \leq Q^2 \leq 65$~GeV$^2$ with a luminosity of 2.0~pb$^{-1}$
collected in 1994.
The event selection criteria have been very similar to the inclusive
$F_2$ analysis adding the requirement of a gap in the 
pseudo-rapidity range $3.0 < \eta < 7.5$ which restricts the 
proton remnant mass to  $M_Y < 1.6~$GeV. The exchanged object
(`pomeron') carries a fraction 
$\xpom$ of the pomeron momentum in the proton 
measured as $\xpom = x \cdot (Q^2 +M_X^2)/Q^2$ where $M_X^2$ is the
mass of the hadronic system produced with the rapidity gap $\Delta
\eta$ to the proton or its dissociation product. In the partonic view
of the $\pom$ there are quarks and gluons with a fraction
$\beta$ of the pomeron momentum,
i.e. $\beta=x/\xpom=Q^2/(Q^2+M_X^2)$. The mass $M_X$ is measured
with the calorimeter cell energies in appropriate combination with
reconstructed tracks and $Q^2$ and $x$
are determined with the $\Sigma$ method.

\begin{figure} [!htb]
 \vspace{-0.7cm}
 \begin{center}      
 \epsfig{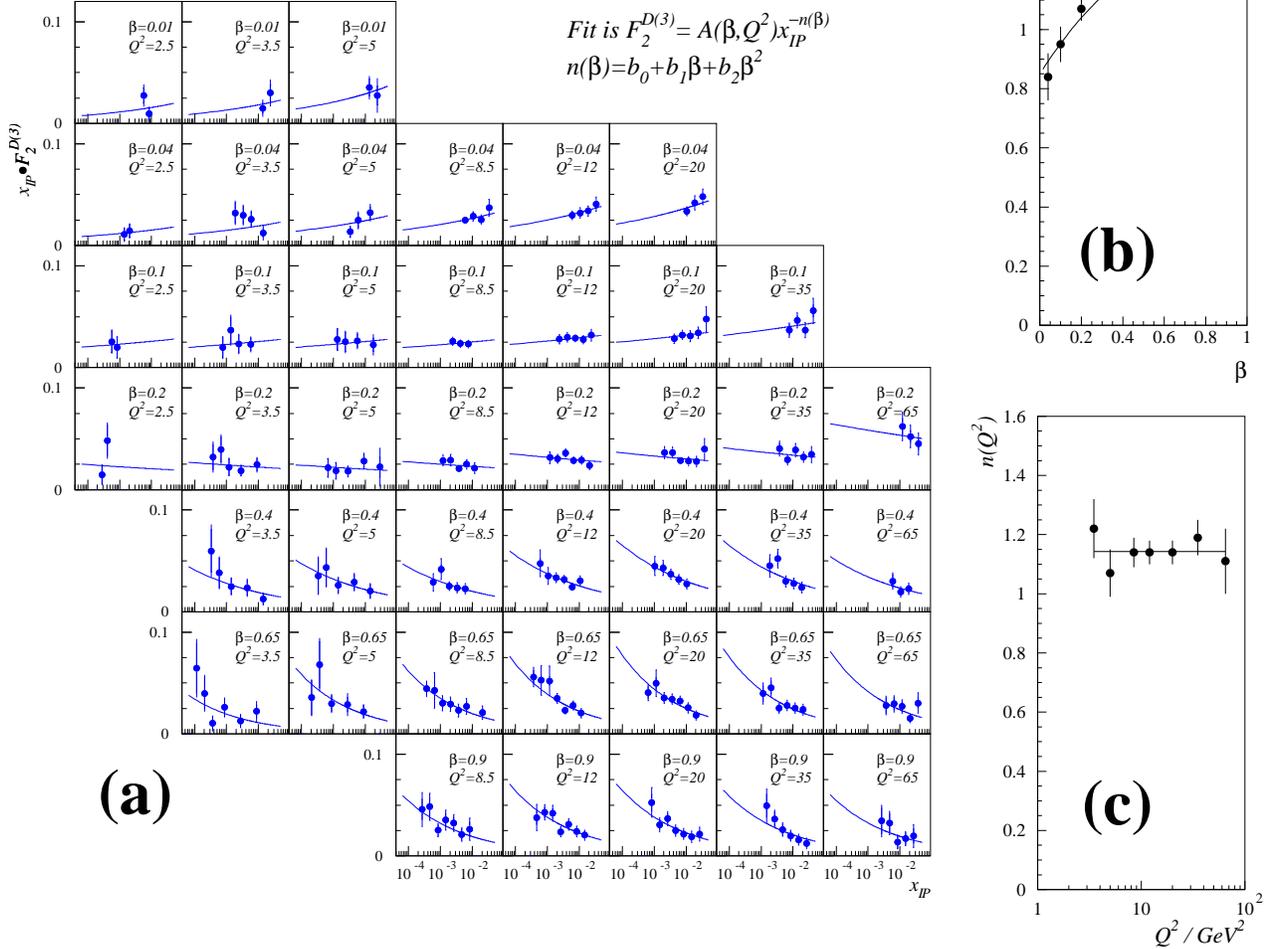}
 \vspace{-0.8cm}
 \caption{(a) \protect{$\xpom \cdot F_2^{D(3)}(\beta,Q^2,\xpom)$} as
   measured by H1 and the  
$\beta$ (b) and $Q^2$ (c) dependence of $n$ determined from fits 
\protect{$F_2^{D(3)}=  A(\beta,Q^2)/ {\xpom}^n$}.
The experimental errors are statistical and systematic added in quadrature.
The  fit described in the text is shown.}
\protect\label{f2d3}   
\end{center}
\vspace{+0.cm}
\end{figure}
Fig.\ref{f2d3} shows the measurement of the structure function~\cite{schlein}
$F_2^{D(3)}(\beta,Q^2,\xpom)$ extracted from the triple differential
cross section divided by the kinematic factor $\kappa$, eq.1, for
$F_L=0$ integrating over $-t < 1$~GeV$^2$ where $t$ is the yet
unmeasured momentum transfer from the proton to the $M_X$ system. The
structure function is well described by a fit $F_2^{D(3)}(\beta,Q^2,\xpom) =
A(\beta, Q^2) \cdot \xpom^{-n}$. This measurement established a
dependence of n on $\beta$ for $\beta \leq 0.3$ which implies that
simple factorization of the deep inelastic diffractive cross section
into a universal flux factor and a structure function $A \propto
F_2^D$ does not hold. This  may be due to an exchange of more than a
single Regge trajectory~\cite{pl}. Apart from the lowest $\beta$ 
the measured $n$ values,
fig.\ref{f2d3}, are still not far from their naive expectation value of $n
= 2 \alpha(0) -1 \simeq 1.17$ introducing the pomeron Regge trajectory
with an intercept of about 1.085 as derived from soft hadron
diffraction experiments.

Integrating $F_2^{3(D)}$ over $\xpom$ from 0.0003 to 0.05,
i.e. considering all
accessed momenta of the pomeron in the proton, H1 obtained the
structure function $\tilde{F}_2^D$ which rises with $Q^2$ up to $\beta$ near
0.9 but does not depend on $\beta$ in the covered $Q^2$ range \cite{paul}.     
This can be explained with gluon dominance in the
pomeron up to largest $\beta$ in an appealing
attempt to quantify the $Q^2, \beta$ behaviour of $\tilde{F}_2^D$ in a DGLAP
evolution approach.
For $Q^2$
between 2 and 70~GeV$^2$ the gluon carries 90 to 80\% of the pomeron
momentum, respectively. This is consistent with the experimental observation of
diffractive $D^*$ production and the study of energy flow in the
$\gamma^* \pom$ cms~\cite{tapp}: central particle production is observed as
expected from a three parton `final' state, e.g.  from two quarks from 
photon-gluon
fusion and a radiated second gluon. A quark object would predominantly
produce two partons giving rise to particle production aligned with
the $\gamma^* \pom$ axis, contrary to what is observed experimentally.
The diffractive structure function measurement is as well consistent
with diffractive jet production \cite{thei} and event shape analyses \cite{valk}. 
The systematic errors of $F_2^{D(3)}$ are still about 20\%
due to energy scale errors and simulation uncertainties. A
precise investigation of diffraction at HERA is still ahead. 

\section{Concluding Remarks}
The results presented by H1 to the 1996 DIS conference represent an 
impressive extension of the physics scope as compared to the previous DIS 
conference \cite{john} with a first rather precise $F_2$ measurement, the 
first $F_L$ and $F_2^c$ determinations, the first NLO gluon distribution 
determined from jets, the proton structure probed with $W$ bosons,
 apparently broken factorization of the deep inelastic diffraction 
cross section and further interesting observations.  It can steadily be  
anticipated how impressive the physics at HERA will be in 2005,
at a machine which by then
represents a 30 years long dream and effort to investigate
with an electron-proton collider
the  substructure of matter and the unification of forces by 
gauge field theories.
It is likely to be precision which at HERA leads
 to new insight requiring
luminosity and patience.
\vspace*{0.4cm} 

$\bf{Acknowledgment}$
The results presented here are due to a huge, collaborative effort of too
many people to be named here. I wish to express special thanks to all
members of the Zeuthen H1 group for their friendly support over the 
lifetime of H1, the H1 structure function group for intense joint work 
over the last years and the Rome speakers of H1 for sharing their
expertise and for  
their generosity when I failed to include their favoured results into this 
presentation. My attempt to summarize the deep inelastic physics of 
H1 would not have been undertaken without the advice and encouragement of 
John Dainton, Albert DeRoeck, Ralph Eichler and Joel Feltesse. Thanks are 
due to Guilio D'Agostini and his team for organizing a remarkable 
conference in Rome, and to Andrea Nigro for a very friendly approach
towards  
completion of this writeup. I finally thank the 
former minister of science and education 
of Italy, our colleague Giorgio Salvini, for making me think harder about
the past and the success of the deep inelastic physics research at HERA. 

¥
\end{document}